\documentclass[aps,pra,reprint,superscriptaddress]{revtex4-1}
\usepackage{graphicx}
\usepackage{amsmath}
\usepackage{amssymb}
\usepackage[colorlinks,linkcolor=blue,
            citecolor=blue,
            hyperindex,
            pdfstartview=FitH,
            plainpages=false]
            {hyperref}
\usepackage{ulem}

\newcommand {\bscco}{Bi$_2$Sr$_2$CaCu$_2$O$_{8+\delta}$}
\newcommand {\uJcm}{$\mu$J/cm$^{2}$}
\newcommand {\nJcm}{nJ/cm$^{2}$}

\begin{document}
\title{Photoinduced changes of the chemical potential in superconducting \bscco}

\author{Tristan L. Miller}
\author{Christopher L. Smallwood}
\author{Wentao Zhang}
\affiliation{Materials Sciences Division, Lawrence Berkeley National Laboratory, Berkeley, California 94720, USA}
\affiliation{Department of Physics, University of California, Berkeley, California 94720, USA}
\author{Hiroshi Eisaki}
\affiliation{Electronics and Photonics Research Institute, National Institute of Advanced Industrial Science and Technology, Ibaraki 305-8568, Japan}
\author{Joseph Orenstein}
\author{Alessandra Lanzara}
\email{alanzara@lbl.gov}
\affiliation{Materials Sciences Division, Lawrence Berkeley National Laboratory, Berkeley, California 94720, USA}
\affiliation{Department of Physics, University of California, Berkeley, California 94720, USA}
\date {\today}


\begin{abstract}

The chemical potential of a superconductor is of critical importance since, at equilibrium, it is the energy where electrons pair and form the superconducting condensate.  However, in non-equilibrium measurements, there may be a difference between the chemical potential of the quasiparticles and that of the pairs.  Here we report a systematic time- and angle-resolved photoemission study of the pump-induced change in the chemical potential of an optimally doped \bscco{} (Bi2212) sample in both its normal and superconducting states.  The change in chemical potential can be understood by separately considering the change in the valence band energy relative to the vacuum, and the change in chemical potential relative to the valence band energy.  We attribute the former effect to a changing potential barrier at the sample surface, and the latter effect to the conservation of charge in an asymmetrical density of states.  The results indicate that the pair and quasiparticle chemical potentials follow each other even on picosecond timescales.

\end{abstract}

\maketitle

\section{Introduction}
\label{introduction}

Although high-temperature superconductors are not fully understood on a microscopic level, the most important activity occurs at the chemical potential energy, where Cooper pairs form the superconducting condensate.  Thus the manipulation of the chemical potential has many basic electronic applications; for example, a difference in the condensate energy across a Josephson junction drives an alternating supercurrent\cite{Josephson1974}.  Ultrafast manipulation of the chemical potential with laser pump pulses has recently been realized by time- and angle-resolved photoemission spectroscopy (tr-ARPES) in several materials\cite{Wang2012,Crepaldi2012,Sobota2012,Crepaldi2013}, including high-temperature superconductors\cite{Avigo2013,Yang2014,Rameau2014,Smallwood2014}.

Transport experiments on superconducting Sn have demonstrated a distinction between the chemical potential of the quasiparticles and the chemical potential of Cooper pairs, reporting that the two equilibrate on a timescale of at least 200 ps\cite{Clarke1972,Tinkham1972}.  Given the subpicosecond timescales so far observed by tr-ARPES on high-temperature superconductors, the question is raised whether the two chemical potentials are in equilibrium.  So far, most studies cannot address this question since they only look at superconducting materials in their normal states\cite{Avigo2013,Yang2014,Rameau2014}.

Studying the chemical potential in the superconducting state is difficult because of the presence of the superconducting energy gap, which represents the energy required to break Cooper pairs.  This energy gap is centered on the pair chemical potential\cite{Tinkham1972}, and studies on cuprate superconductor \bscco{} (Bi2212) show that the gap's size changes upon pumping\cite{Smallwood2012,Smallwood2014,Zhang2014}.  Since only the lower edge of the gap is observed, it cannot be used to accurately measure the chemical potential. However, the well-known \textit{d}-wave symmetry of cuprate superconductors offers a way forward because the gap vanishes along the (0,0)--($\pi$,$\pi$) momentum direction.  Along this direction, the quasiparticle chemical potential, if not the pair chemical potential, can be determined by the energy distribution of quasiparticles.

Here we perform a systematic study of the pump-induced shift in the electron energies of optimally doped Bi2212, in the superconducting and normal states.  We measure the change in the quasiparticle chemical potential relative to vacuum ($\Delta\mu_{vac}$) and the change in the valence band energy relative to vacuum ($\Delta\varepsilon$).  Both quantities relax exponentially, with a timescale of 2 ps in the superconducting state and $<1$ ps in the normal state.  The quantity $\Delta\varepsilon$ is consistent with a pump-induced change in the potential barrier at the sample surface.

In an analysis of the pump-induced change in quasiparticle chemical potential relative to the valence band energy ($\Delta\mu_\varepsilon$), we find that $\Delta\mu_\varepsilon$ is negative in the superconducting state and positive in the normal state, and that the energy gap is centered at the chemical potential.  We use a simple model to explain these observations in terms of conservation of charge in an asymmetrical density of states, and find that the model best fits the data if the energy gap is centered at the chemical potential.  We conclude that the pair and quasiparticle chemical potentials of Bi2212 follow each other in even as they change on a picosecond timescale.

\section{Experimental Details}
\label{experimental details}

The tr-ARPES setup operates  with 836 nm pump pulses ($h\nu=1.48$ eV) and 209 nm probe pulses ($h\nu=5.93$ eV) (see Ref. \cite{Smallwood2012RSI} for a detailed description).  The spot size is about 100 $\mu$m for the pump and 50 $\mu$m for the probe. The energy and momentum resolutions are 23 meV, 0.003 \AA{}$^{-1}$ respectively.  The time resolution is 300 fs, longer than the 100 fs it takes for electrons to reach a quasithermal distribution\cite{Perfetti2007,Graf2011}, so the chemical potential is well-defined at all times.

Since this study measures energy very precisely, particular care was taken to remove systematic errors in the energy.  Corrections are made to take into account the nonlinear response of the detector\cite{Smallwood2012RSI,Reber2014}.  To distinguish changes in the chemical potential from the suppression of the gap, all measurements in this study were taken along the nodal direction ($\Gamma$--Y) of an optimally doped Bi2212 sample. The alignment is sufficiently precise that the gap size along this cut can be no more than 0.7 meV.   In all measurements, the probe fluence is 3 \nJcm{} (photoemitting $10^6$ electrons/cm$^2$/pulse); significantly higher probe fluences do not reproduce the same results because the mirror charge effect\cite{Zhou2005,Graf2010} becomes significant, and is modified upon pumping.  Because small instabilities in the laser power are linked to the space charge and mirror charge effects, the electron energies also drift over time.  Therefore, each measurement is taken over many repeated cycles, and a linear correction is made in the electron energy vs. time (no more than 0.2 meV per cycle).

\section{Results}
\label{results}

\subsection{Chemical Potential and Dispersion energy}
\label{resultsA}

\begin{figure}\centering\includegraphics[width=3.375in]{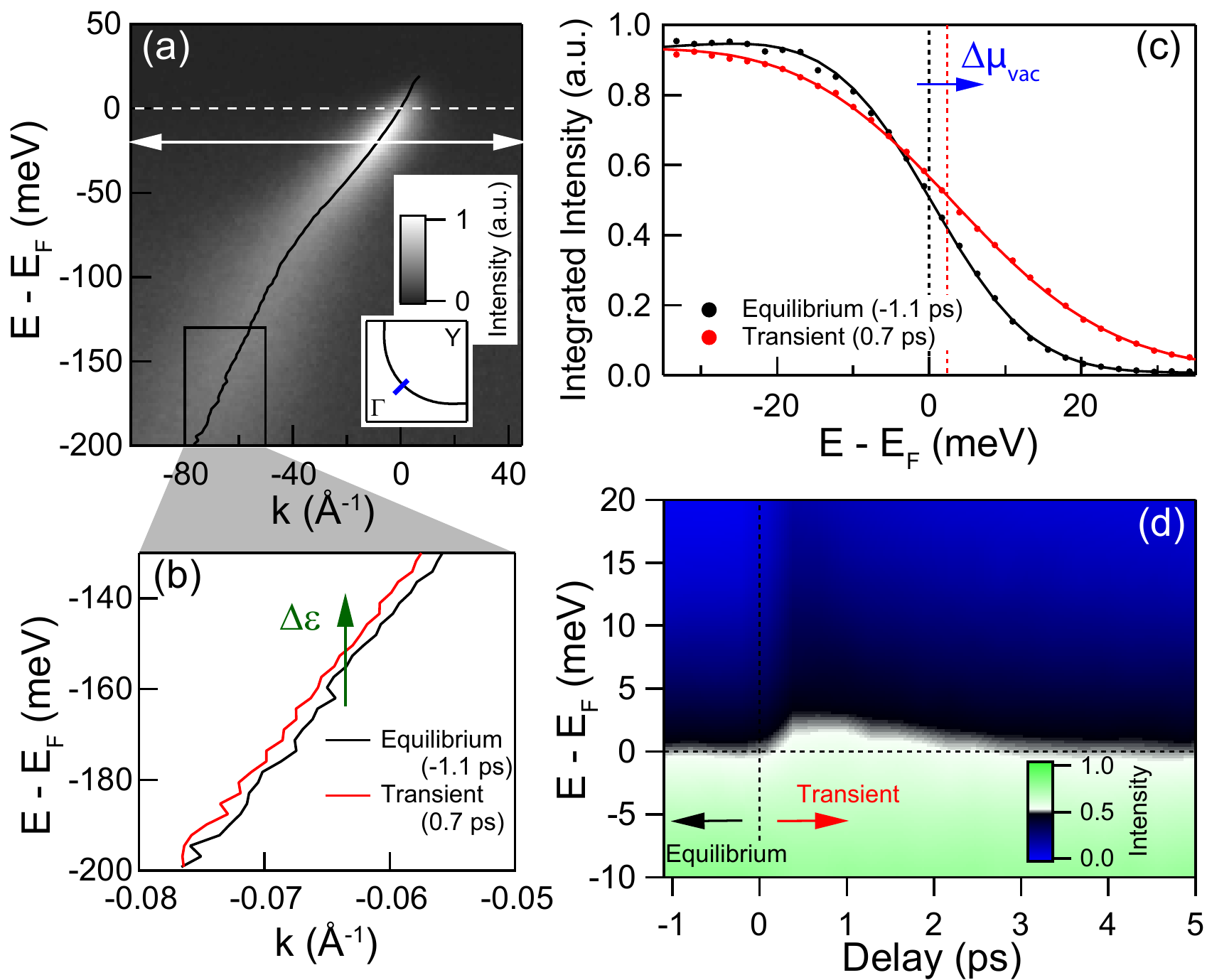}
\caption{(a) The ARPES intensity map of optimally doped Bi2212 at 35 K along the cut shown in the inset. (b) The valence dispersion of the sample in (a), between 130 and 200 meV binding energy, both before and after the arrival of a 20 \uJcm{} pump pulse.  (c) The ARPES intensity integrated along the momentum range indicated by the white double arrow in (a), both before and after the pump pulse.  Solid lines show fits to Fermi-Dirac functions, and dashed lines show the chemical potential from the fits. (d) The integrated ARPES intensity as a function of the delay time between the pump and probe.
}
\label{Fig1}
\end{figure}

In our study, we define two distinct ways to look at changes in the quasiparticle energies.  First, we look at the valence band dispersion at a given momentum, and measure the change in its energy relative to the vacuum level ($\Delta\varepsilon$).  Second, we look at the quasiparticle chemical potential relative to the vacuum ($\Delta\mu_{vac}$), as determined by the Fermi-Dirac distribution of quasiparticles.

Each of these changes is illustrated with tr-ARPES data in Figure \ref{Fig1}.  Panel (a) shows the ARPES intensity map of an optimally doped Bi2212 sample (T$_c$ = 91K) at 35 K.  The quasiparticle dispersion (black curve) is extracted using the standard method of fitting horizontal cuts of the intensity to Lorentzian curves\cite{LaShell2000}.  The sample is pumped with a 20 \uJcm{} laser pulse, and the transient quasiparticle dispersion (0.7 ps delay after the pump pulse) is compared with the equilibrium dispersion [see Fig. \ref{Fig1}(b)].  Data show a clear transient shift in the dispersion of about 5 meV.  Note that we only compare the dispersion in the range of 130--200 meV binding energy, since pumping is known to affect the electron-boson coupling renormalization near 70 meV\cite{Rameau2014,Zhang2014}, and near the Fermi level there is a systematic bias from the instrumental energy resolution\cite{Levy2014}.

Figure \ref{Fig1}(c) shows the integrated ARPES intensity over a small momentum range [double arrow in panel (a)].  Using the same method as Ref. \cite{Smallwood2014}, we fit the equilibrium and transient intensity to a Fermi-Dirac distribution convolved with a Gaussian for the energy resolution.  The chemical potential $\mu$ clearly increases by about 2.5 meV in response to the pump.  In Figure \ref{Fig1}(d) we show the integrated intensity as a function of delay time, visually illustrating that $\mu$ increases and then relaxes over a few picoseconds.

\begin{figure}\centering\includegraphics[width=3.375in]{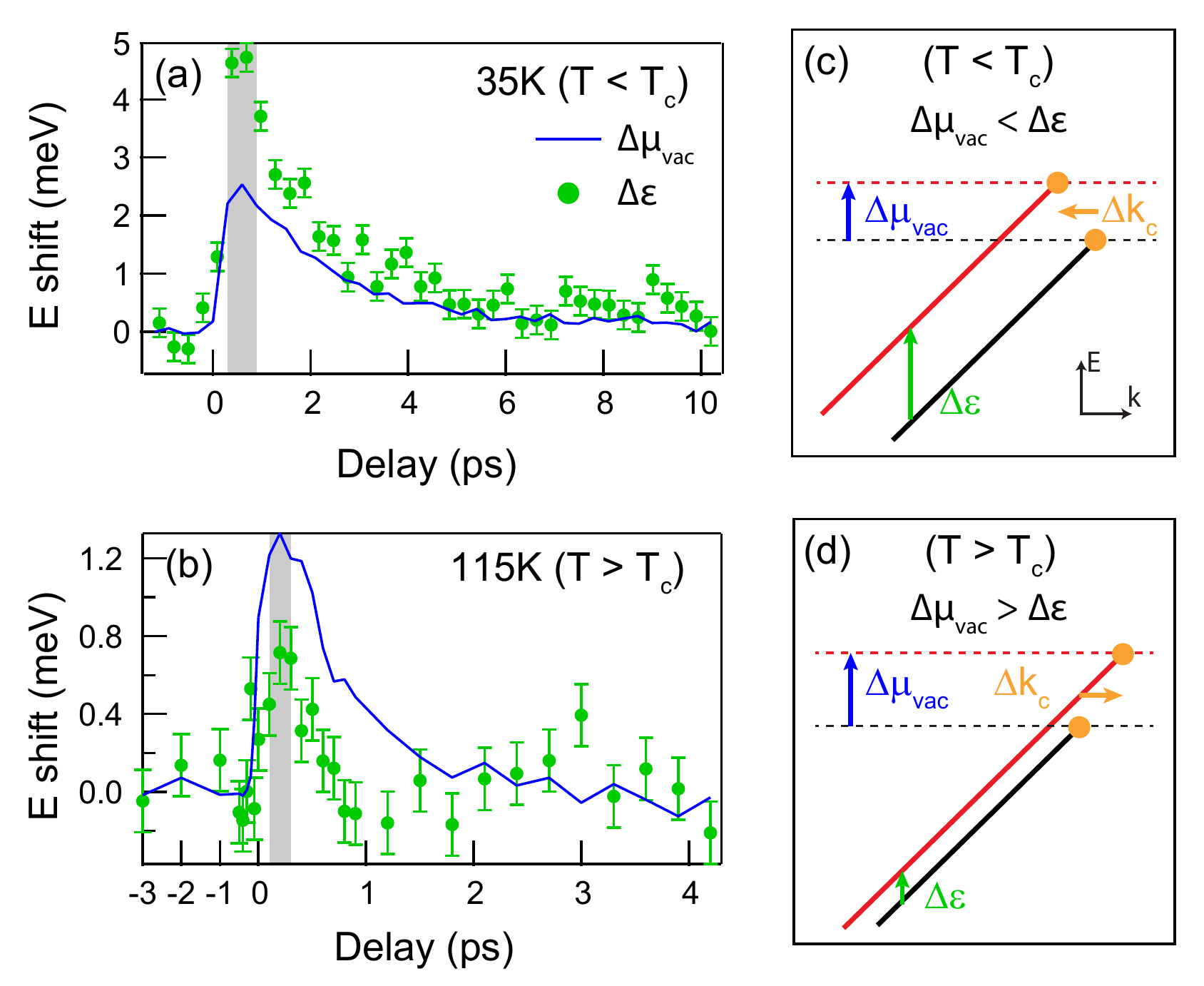}
\caption{The change in chemical potential ($\Delta\mu_{vac}$) and the shift in quasiparticle dispersion ($\Delta\varepsilon$) relative to the vacuum energy in response to a 20 \uJcm{} pump pulse at 35 K (a) and 115 K (b).  (c,d) Illustrations of the changes shown in (a,b).  $k_c$ is the point where the dispersion crosses the chemical potential.}
\label{Fig2}
\end{figure}

Figure \ref{Fig2} shows $\Delta\mu_{vac}$ and $\Delta\varepsilon$ together as a function of delay time in the superconducting [panel (a)] and normal [panel (b)] states of optimally doped Bi2212.  Although both quantities increase upon pumping and recover exponentially with a timescale of $\sim$2 ps in the superconducting state ($<$1 ps in the normal state), it is important to note the difference in their magnitudes.  We define $\Delta\mu_\varepsilon$ as the change in chemical potential relative to the valence band energy, that is, $\Delta\mu_\varepsilon = \Delta\mu_{vac} - \Delta\varepsilon$.  In the superconducting state, $\Delta\mu_\varepsilon$ is negative, while in the normal state, it is positive.  This is schematically illustrated in Fig. \ref{Fig2}(c) and \ref{Fig2}(d).

The quantity $\Delta\mu_{vac}$ is the sum of two effects: $\Delta\mu_\varepsilon$ and $\Delta\varepsilon$.  $\Delta\varepsilon$ is a uniform change in the energy of all photoemitted electrons, which can be economically explained by a pump-induced change in the potential energy barrier at the surface of the sample.  Photoemitted electrons cross this surface barrier in a matter of femtoseconds, because the mean free electron path of 6 eV electrons in a solid is only $\sim$5 nm\cite{Seah1979}.  $\Delta\mu_\varepsilon$, on the other hand, signifies a change in the occupation of available quasiparticle states.  A possible mechanism for this change will be discussed in section \ref{discussion}.

\subsection{Pump Fluence Dependence}
\label{resultsB}

\begin{figure}\centering\includegraphics[width=3.375in]{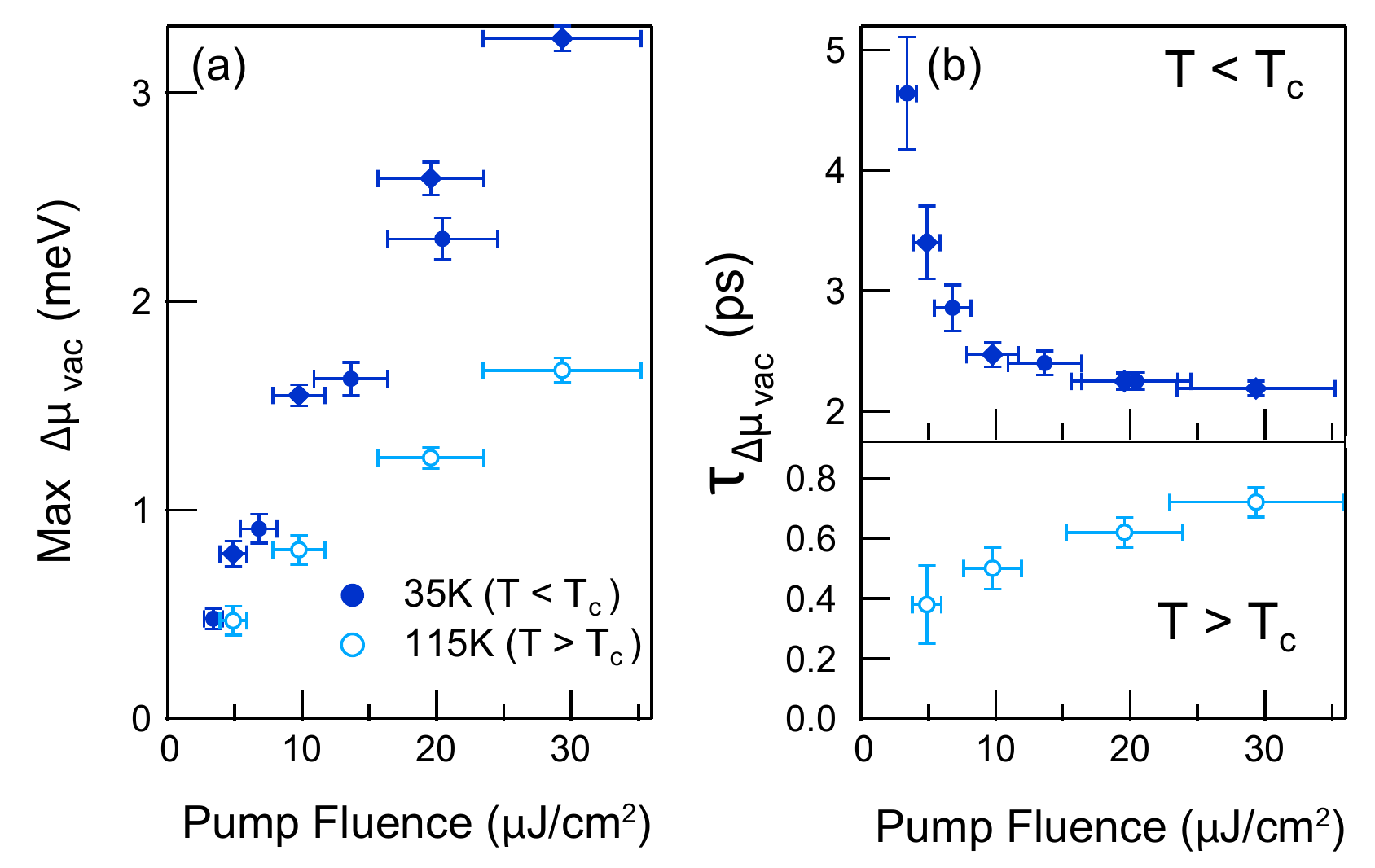}
\caption{(a) The maximum shift $\Delta\mu_{vac}$, determined by averaging over the gray bars in Figure \ref{Fig2}(a,b).  (b) The recovery rates of $\mu$, as determined by simple exponential fitting after 0.6 ps.}
\label{Fig3}
\end{figure}

We turn our focus to $\Delta\mu_{vac}$, performing a systematic study of delay time, temperature, and pump fluence in Figure \ref{Fig3}.  Panel (a) shows the maximum $\Delta\mu_{vac}$ in the superconducting and normal states as a function of pump fluence.  $\Delta\mu_{vac}$ is observed to be larger in the superconducting state, and to increase with pump fluence.  Panel (b) shows the recovery time of $\Delta\mu_{vac}$, as determined by a simple exponential fit from 0.6 ps delay time onwards.  In the superconducting state, the recovery time becomes shorter at greater pump fluences, saturating at about 2 ps.  In the normal state, the recovery is less than 1 ps, and becomes longer at greater pump fluences.

The timescale of the change in chemical potential can be compared to many other changes occurring simultaneously in the system.  With a 22 \uJcm{} pump pulse, the near-nodal superconducting gap is closed and recovers with an initial timescale of 2 ps, although this rate slows down after a few picoseconds\cite{Smallwood2014}.  In the superconducting state, similar fluences suppress the nodal electron-boson renormalization, allowing it to recover on a 4 ps timescale\cite{Zhang2014}.  But most suggestively, the chemical potential dynamics are qualitatively similar to the nodal quasiparticle population.  The nodal quasiparticle relaxation rate increases with pump fluence because of bimolecular recombination, and is much faster in the normal state because excited quasiparticles can combine with thermal quasiparticles\cite{Smallwood2012,Smallwood2015}.  The trends seen in the chemical potential dynamics may be explained in the same way.  In the superconducting state, the timescales of the nodal quasiparticles are slower overall than those of the chemical potential, but this may merely indicate that the relationship between the chemical potential and quasiparticle population is nonlinear.

\section{Discussion}
\label{discussion}

\begin{figure*}\centering\includegraphics[width=7in]{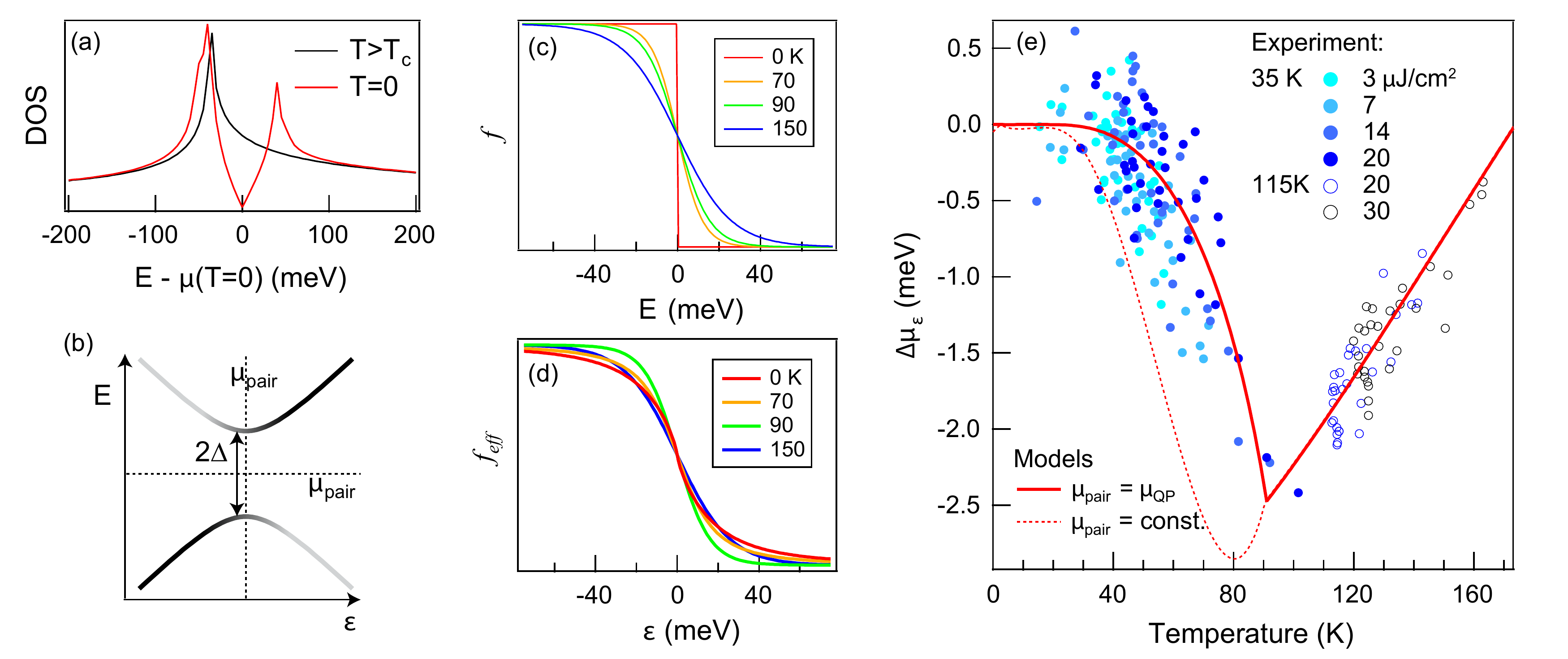}
\caption{(a) The density of states of optimally doped Bi2212 calculated with a tight-binding model, in both the normal state, and superconducting state at T=0.  (b) An illustration of the relationship between $\varepsilon$, the quasiparticle energies before the pairing interaction is turned on, and $E$, the energies afterwards. (c) The distribution function as a function of E. (d) The effective distribution function as a function of $\varepsilon$. (e) A comparison of the experimental $\Delta\mu_\varepsilon$ to its calculated value, considering the case where the pair chemical potential is equal to the quasiparticle chemical potential and the case where it is held constant.
}
\label{Fig4}
\end{figure*}

The most surprising result of this study is that the chemical potential relative to the valence band energy increases in the normal state, but decreases in the superconducting state.  This cannot be explained by a change in charge density, because electrons cannot travel very far from the measurement area on a picosecond timescale, and the number of electrons photoemitted from the sample is unaffected by the pump pulse.  Assuming conservation of charge, we find the chemical potential using
\begin{equation}
 N = \int_{-\infty}^{\infty} D(E) f(E-\mu_{QP},T) \mathrm{d}E \label{charge conservation},
\end{equation}
where $D(E)$ is the density of states at energy $E$, and $f(E,T)$ is the Fermi-Dirac distribution function at electronic temperature $T$ (which need not match the temperature of the rest of the system), and $N$ is the density of electrons.  As is well-known in metals, the quasiparticle chemical potential $\mu_{QP}$ must change as a function of $D(E)$ and $T$ in order to conserve charge.  The distinction we draw between $\Delta\varepsilon$ and $\Delta\mu_\varepsilon$ is well justified in this picture, since the former isolates the contribution to $\Delta\mu_{vac}$ from uniform shifts in $D(E)$, while the latter isolates the contribution from the change in $T$.

Before introducing superconductivity, the density of states can be approximated with a simple phenomenological tight-binding model\cite{Norman1995} [see Fig. \ref{Fig4}(a)].  Because of the saddle point in the valence band, there is a Van Hove singularity in the density of states just below the chemical potential, which causes an asymmetric density of states.  An asymmetric density of states could also have other causes, such as particle-hole asymmetry in the pseudogap\cite{Yang2008,Hashimoto2010,Sakai2013,Rice2012}, but we do not include such effects in this basic model.  The asymmetry in the density of states near the chemical potential would cause a positive $\Delta\mu_\varepsilon$ as $T$ increases.  Since pumping increases $T$ in all our measurements, this is sufficient to explain the pump-induced increase of $\Delta\mu_\varepsilon$ in the normal state.

In the superconducting state, the pairing interaction causes the electron and hole quasiparticles to mix.  Below, we explain a sense in which the distribution function $f(E,T)$ is broadened by the pairing interaction.  This broadening causes $\mu_\varepsilon$ to be higher in the superconducting state.  Pumping is known to suppress or destroy superconductivity\cite{Smallwood2012,Smallwood2014,Zhang2014,Smallwood2015}, which should cause a negative $\Delta\mu_\varepsilon$, exactly as we observe.

More precisely, it is necessary to include the effect of the superconducting gap on the density of states.  We distinguish between $\varepsilon$, the quasiparticle energy before the pairing interaction is turned on, and $E$, the energy afterwards, each measured relative to $\mu_{pair}$, the pair chemical potential.  Figure \ref{Fig4}(b) shows the well-known Bogoliubov quasiparticle dispersion\cite{Tinkham1996}, with $E_\pm(\varepsilon) = \pm\sqrt{\varepsilon^2+\Delta^2}$ as the energy of each branch, and $w_\pm(\varepsilon) = \frac{1}{2} (1 + \frac{\varepsilon}{E_\pm} )$ as the spectral weight of each branch.  The resulting density of states is
\begin{align*}
D&(E,\Delta) = \\
&\left\{
\begin{matrix}
\frac{E}{\varepsilon}\Big(w_+(-\varepsilon)D(-\varepsilon) + w_+(\varepsilon)D(\varepsilon)\Big)&\text{for}~ E > \Delta   \\
\frac{E}{\varepsilon}\Big(w_-(-\varepsilon)D(-\varepsilon) + w_-(\varepsilon)D(\varepsilon)\Big)&\text{for}~ E < -\Delta    \\
0&\text{for}~|E| < \Delta
\end{matrix}\right.\tag{2}\label{dos}
\end{align*}
where $\varepsilon = \sqrt{E^2-\Delta^2}$.  In the case of Bi2212, the gap is not constant as a function of momentum because of the \textit{d}-wave gap symmetry, but Figure \ref{Fig4}(a) shows the approximation of summing over a range of gap sizes $\Delta = \Delta_{max}|cos(\Theta)|$, where $\Delta_{max}$ is the maximum gap size, and $\Theta$ is an angle along the Fermi surface.

We now consider the special case where the pair chemical potential and quasiparticle chemical potentials are equal.  Using Equation (\ref{dos}), it is possible to change variables in the integral of Eq. (\ref{charge conservation}) from $E$ to $\varepsilon$.  The new equation is
\begin{equation}
 N = \int_{-\infty}^{\infty} D(\varepsilon) f_{e\!f\!f}(\varepsilon-\mu,T,\Delta) \mathrm{d}\varepsilon \tag{3}\label{fEff definition},
\end{equation}
where $f_{e\!f\!f}$ is the effective distribution function with respect to $\varepsilon$.  When the gap is constant, $f_{e\!f\!f}$ is given exactly by
\begin{equation}
f_{e\!f\!f}(\varepsilon,T,\Delta) = \frac{1}{2} \Big( 1 - \frac{\varepsilon}{E} \Big) + \frac{\varepsilon}{E} f\Big(E,T\Big) \tag{4}\label{fswave},
\end{equation}
which is equivalent to the equation used by Ref. \cite{Marel1990}.  The first term of Eq. (\ref{fswave}) is the contribution from the energy gap, while the second term is the contribution from temperature.  With \textit{d}-wave gap symmetry it is necessary to average Eq. (\ref{fswave}) over a range of gap sizes, but the qualitative characteristics remain the same.

The functions $f(E,T)$ and $f_{e\!f\!f}(\varepsilon,T,\Delta)$ are shown in Fig. \ref{Fig4}(c) and \ref{Fig4}(d), using a \textit{d}-wave gap that follows a BCS temperature dependence and has a maximum size of 39 meV\cite{Vishik2012}.  While $f(E,T)$ gets monotonically broader as T increases, $f_{e\!f\!f}(\varepsilon,T,\Delta)$ actually gets narrower as T increases to $T_c = 91 K$ because the energy gap is closing. The model predicts that $\Delta\mu_\varepsilon$ follows the same trend as the width of $f_{e\!f\!f}$, as shown by the solid red line of Figure \ref{Fig4}(e).  On the same plot, the model is compared to experimental measurements of $\Delta\mu_\varepsilon$ and electronic temperatures over many delay times and fluences.  Despite the lack of free parameters in the model, it matches the data quite well.  We also find that the data does not match the alternate model (dashed red line) in which the pair chemical potential remains constant.  This indicates that instead of remaining constant, the pair chemical potential tracks the quasiparticle chemical potential throughout the experiment.

This simple model considers a homogeneous system with constant doping, but it may be complicated by known inhomogeneities in \bscco{} on length scales of 1--3 nm\cite{Pan2001,Lang2002}.  There may be different domains with different energy gaps, different chemical potentials, and different dynamics.  Since the probe spot size (50 $\mu$m) is much larger than the length scale of inhomogeneities, the apparent chemical potential will be an average of that of the different domains.  Further study is needed to investigate the consequences of a more complex model.

Two other effects were considered and rejected as explanations for the observed chemical potential shifts.  If the crystal lattice should expand, this will cause the entire Brillouin zone to shrink, shifting the dispersion.  However, even with the generous assumption that the lattice is at equilibrium with the electronic temperature, the effect is negligible in our experiments.  For example, in the measurement in Figure \ref{Fig2}(a), the lattice constant should expand by 0.021\%{}\cite{Mouallem-Bahout1994,Asahi1997}, corresponding to a $\Delta\varepsilon$ of only 0.14 meV.  Another effect comes from electric fields in space implied by the change in the potential barrier at the surface.  The length scale of these electric fields is that of the pump spot size ($\sim 100 \mu$m), but this is much larger than the distance the electron travels over 2 ps ($\sim 1 \mu$m).  Therefore, the electric fields have a negligible effect.


\section{Conclusion}
\label{conclusion}

By careful study of the laser-induced change in quasiparticle chemical potential of optimally doped Bi2212 we have shown that the pair chemical potential changes along with it on a picosecond timescale.  The pair chemical potential is significant to any electronic applications, and even a small difference of 1 meV across a Josephson junction would drive an alternating supercurrent with an oscillation period of 2 ps\cite{Josephson1974}, which is incidentally similar to the timescale of the change in chemical potential.

Given the successful use of the chemical potential to study the superconducting gap, it may be fruitful to apply the same technique to the pseudogap state of cuprate superconductors.  The nature of the pseudogap is not well understood, and there is a distinct possibility that, unlike the superconducting gap, the pseudogap is not centered at the chemical potential\cite{Yang2008,Hashimoto2010,Sakai2013,Rice2012}.  Such a particle-hole asymmetry would have an observable effect in the pump-induced shift of the chemical potential in underdoped cuprates.

\begin{acknowledgments}
We thank J. Clarke and D.-H. Lee for helpful discussions.  This work was supported by Berkeley Lab's program on Ultrafast Materials, funded by the U.S. Department of Energy, Office of Science, Office of Basic Energy Sciences, Materials Sciences and Engineering Division, under Contract No. DE-AC02-05CH11231.
\end{acknowledgments}

\bibliographystyle{apsrev4-1}
\bibliography{C:/Users/Tristan/Desktop/JabRef/tlmbib}

\begin{thebibliography}{35}%
\makeatletter
\providecommand \@ifxundefined [1]{%
 \@ifx{#1\undefined}
}%
\providecommand \@ifnum [1]{%
 \ifnum #1\expandafter \@firstoftwo
 \else \expandafter \@secondoftwo
 \fi
}%
\providecommand \@ifx [1]{%
 \ifx #1\expandafter \@firstoftwo
 \else \expandafter \@secondoftwo
 \fi
}%
\providecommand \natexlab [1]{#1}%
\providecommand \enquote  [1]{``#1''}%
\providecommand \bibnamefont  [1]{#1}%
\providecommand \bibfnamefont [1]{#1}%
\providecommand \citenamefont [1]{#1}%
\providecommand \href@noop [0]{\@secondoftwo}%
\providecommand \href [0]{\begingroup \@sanitize@url \@href}%
\providecommand \@href[1]{\@@startlink{#1}\@@href}%
\providecommand \@@href[1]{\endgroup#1\@@endlink}%
\providecommand \@sanitize@url [0]{\catcode `\\12\catcode `\$12\catcode
  `\&12\catcode `\#12\catcode `\^12\catcode `\_12\catcode `\%12\relax}%
\providecommand \@@startlink[1]{}%
\providecommand \@@endlink[0]{}%
\providecommand \url  [0]{\begingroup\@sanitize@url \@url }%
\providecommand \@url [1]{\endgroup\@href {#1}{\urlprefix }}%
\providecommand \urlprefix  [0]{URL }%
\providecommand \Eprint [0]{\href }%
\providecommand \doibase [0]{http://dx.doi.org/}%
\providecommand \selectlanguage [0]{\@gobble}%
\providecommand \bibinfo  [0]{\@secondoftwo}%
\providecommand \bibfield  [0]{\@secondoftwo}%
\providecommand \translation [1]{[#1]}%
\providecommand \BibitemOpen [0]{}%
\providecommand \bibitemStop [0]{}%
\providecommand \bibitemNoStop [0]{.\EOS\space}%
\providecommand \EOS [0]{\spacefactor3000\relax}%
\providecommand \BibitemShut  [1]{\csname bibitem#1\endcsname}%
\let\auto@bib@innerbib\@empty
\bibitem [{\citenamefont {Josephson}(1974)}]{Josephson1974}%
  \BibitemOpen
  \bibfield  {author} {\bibinfo {author} {\bibfnamefont {B.~D.}\ \bibnamefont
  {Josephson}},\ }\href {\doibase 10.1103/RevModPhys.46.251} {\bibfield
  {journal} {\bibinfo  {journal} {Rev. Mod. Phys.}\ }\textbf {\bibinfo {volume}
  {46}},\ \bibinfo {pages} {251} (\bibinfo {year} {1974})}\BibitemShut
  {NoStop}%
\bibitem [{\citenamefont {Wang}\ \emph {et~al.}(2012)\citenamefont {Wang},
  \citenamefont {Hsieh}, \citenamefont {Sie}, \citenamefont {Steinberg},
  \citenamefont {Gardner}, \citenamefont {Lee}, \citenamefont
  {Jarillo-Herrero},\ and\ \citenamefont {Gedik}}]{Wang2012}%
  \BibitemOpen
  \bibfield  {author} {\bibinfo {author} {\bibfnamefont {Y.~H.}\ \bibnamefont
  {Wang}}, \bibinfo {author} {\bibfnamefont {D.}~\bibnamefont {Hsieh}},
  \bibinfo {author} {\bibfnamefont {E.~J.}\ \bibnamefont {Sie}}, \bibinfo
  {author} {\bibfnamefont {H.}~\bibnamefont {Steinberg}}, \bibinfo {author}
  {\bibfnamefont {D.~R.}\ \bibnamefont {Gardner}}, \bibinfo {author}
  {\bibfnamefont {Y.~S.}\ \bibnamefont {Lee}}, \bibinfo {author} {\bibfnamefont
  {P.}~\bibnamefont {Jarillo-Herrero}}, \ and\ \bibinfo {author} {\bibfnamefont
  {N.}~\bibnamefont {Gedik}},\ }\href {\doibase 10.1103/PhysRevLett.109.127401}
  {\bibfield  {journal} {\bibinfo  {journal} {Phys. Rev. Lett.}\ }\textbf
  {\bibinfo {volume} {109}},\ \bibinfo {pages} {127401} (\bibinfo {year}
  {2012})}\BibitemShut {NoStop}%
\bibitem [{\citenamefont {Crepaldi}\ \emph {et~al.}(2012)\citenamefont
  {Crepaldi}, \citenamefont {Ressel}, \citenamefont {Cilento}, \citenamefont
  {Zacchigna}, \citenamefont {Grazioli}, \citenamefont {Berger}, \citenamefont
  {Bugnon}, \citenamefont {Kern}, \citenamefont {Grioni},\ and\ \citenamefont
  {Parmigiani}}]{Crepaldi2012}%
  \BibitemOpen
  \bibfield  {author} {\bibinfo {author} {\bibfnamefont {A.}~\bibnamefont
  {Crepaldi}}, \bibinfo {author} {\bibfnamefont {B.}~\bibnamefont {Ressel}},
  \bibinfo {author} {\bibfnamefont {F.}~\bibnamefont {Cilento}}, \bibinfo
  {author} {\bibfnamefont {M.}~\bibnamefont {Zacchigna}}, \bibinfo {author}
  {\bibfnamefont {C.}~\bibnamefont {Grazioli}}, \bibinfo {author}
  {\bibfnamefont {H.}~\bibnamefont {Berger}}, \bibinfo {author} {\bibfnamefont
  {P.}~\bibnamefont {Bugnon}}, \bibinfo {author} {\bibfnamefont
  {K.}~\bibnamefont {Kern}}, \bibinfo {author} {\bibfnamefont {M.}~\bibnamefont
  {Grioni}}, \ and\ \bibinfo {author} {\bibfnamefont {F.}~\bibnamefont
  {Parmigiani}},\ }\href {\doibase 10.1103/PhysRevB.86.205133} {\bibfield
  {journal} {\bibinfo  {journal} {Phys. Rev. B}\ }\textbf {\bibinfo {volume}
  {86}},\ \bibinfo {pages} {205133} (\bibinfo {year} {2012})}\BibitemShut
  {NoStop}%
\bibitem [{\citenamefont {Sobota}\ \emph {et~al.}(2012)\citenamefont {Sobota},
  \citenamefont {Yang}, \citenamefont {Analytis}, \citenamefont {Chen},
  \citenamefont {Fisher}, \citenamefont {Kirchmann},\ and\ \citenamefont
  {Shen}}]{Sobota2012}%
  \BibitemOpen
  \bibfield  {author} {\bibinfo {author} {\bibfnamefont {J.~A.}\ \bibnamefont
  {Sobota}}, \bibinfo {author} {\bibfnamefont {S.}~\bibnamefont {Yang}},
  \bibinfo {author} {\bibfnamefont {J.~G.}\ \bibnamefont {Analytis}}, \bibinfo
  {author} {\bibfnamefont {Y.~L.}\ \bibnamefont {Chen}}, \bibinfo {author}
  {\bibfnamefont {I.~R.}\ \bibnamefont {Fisher}}, \bibinfo {author}
  {\bibfnamefont {P.~S.}\ \bibnamefont {Kirchmann}}, \ and\ \bibinfo {author}
  {\bibfnamefont {Z.-X.}\ \bibnamefont {Shen}},\ }\href {\doibase
  10.1103/PhysRevLett.108.117403} {\bibfield  {journal} {\bibinfo  {journal}
  {Phys. Rev. Lett.}\ }\textbf {\bibinfo {volume} {108}},\ \bibinfo {pages}
  {117403} (\bibinfo {year} {2012})}\BibitemShut {NoStop}%
\bibitem [{\citenamefont {Crepaldi}\ \emph {et~al.}(2013)\citenamefont
  {Crepaldi}, \citenamefont {Cilento}, \citenamefont {Ressel}, \citenamefont
  {Cacho}, \citenamefont {Johannsen}, \citenamefont {Zacchigna}, \citenamefont
  {Berger}, \citenamefont {Bugnon}, \citenamefont {Grazioli}, \citenamefont
  {Turcu}, \citenamefont {Springate}, \citenamefont {Kern}, \citenamefont
  {Grioni},\ and\ \citenamefont {Parmigiani}}]{Crepaldi2013}%
  \BibitemOpen
  \bibfield  {author} {\bibinfo {author} {\bibfnamefont {A.}~\bibnamefont
  {Crepaldi}}, \bibinfo {author} {\bibfnamefont {F.}~\bibnamefont {Cilento}},
  \bibinfo {author} {\bibfnamefont {B.}~\bibnamefont {Ressel}}, \bibinfo
  {author} {\bibfnamefont {C.}~\bibnamefont {Cacho}}, \bibinfo {author}
  {\bibfnamefont {J.~C.}\ \bibnamefont {Johannsen}}, \bibinfo {author}
  {\bibfnamefont {M.}~\bibnamefont {Zacchigna}}, \bibinfo {author}
  {\bibfnamefont {H.}~\bibnamefont {Berger}}, \bibinfo {author} {\bibfnamefont
  {P.}~\bibnamefont {Bugnon}}, \bibinfo {author} {\bibfnamefont
  {C.}~\bibnamefont {Grazioli}}, \bibinfo {author} {\bibfnamefont {I.~C.~E.}\
  \bibnamefont {Turcu}}, \bibinfo {author} {\bibfnamefont {E.}~\bibnamefont
  {Springate}}, \bibinfo {author} {\bibfnamefont {K.}~\bibnamefont {Kern}},
  \bibinfo {author} {\bibfnamefont {M.}~\bibnamefont {Grioni}}, \ and\ \bibinfo
  {author} {\bibfnamefont {F.}~\bibnamefont {Parmigiani}},\ }\href {\doibase
  10.1103/PhysRevB.88.121404} {\bibfield  {journal} {\bibinfo  {journal} {Phys.
  Rev. B}\ }\textbf {\bibinfo {volume} {88}},\ \bibinfo {pages} {121404}
  (\bibinfo {year} {2013})}\BibitemShut {NoStop}%
\bibitem [{\citenamefont {Avigo}\ \emph {et~al.}(2013)\citenamefont {Avigo},
  \citenamefont {Cort\'es}, \citenamefont {Rettig}, \citenamefont
  {Thirupathaiah}, \citenamefont {Jeevan}, \citenamefont {Gegenwart},
  \citenamefont {Wolf}, \citenamefont {Ligges}, \citenamefont {Wolf},
  \citenamefont {Fink},\ and\ \citenamefont {Bovensiepen}}]{Avigo2013}%
  \BibitemOpen
  \bibfield  {author} {\bibinfo {author} {\bibfnamefont {I.}~\bibnamefont
  {Avigo}}, \bibinfo {author} {\bibfnamefont {R.}~\bibnamefont {Cort\'es}},
  \bibinfo {author} {\bibfnamefont {L.}~\bibnamefont {Rettig}}, \bibinfo
  {author} {\bibfnamefont {S.}~\bibnamefont {Thirupathaiah}}, \bibinfo {author}
  {\bibfnamefont {H.~S.}\ \bibnamefont {Jeevan}}, \bibinfo {author}
  {\bibfnamefont {P.}~\bibnamefont {Gegenwart}}, \bibinfo {author}
  {\bibfnamefont {T.}~\bibnamefont {Wolf}}, \bibinfo {author} {\bibfnamefont
  {M.}~\bibnamefont {Ligges}}, \bibinfo {author} {\bibfnamefont
  {M.}~\bibnamefont {Wolf}}, \bibinfo {author} {\bibfnamefont {J.}~\bibnamefont
  {Fink}}, \ and\ \bibinfo {author} {\bibfnamefont {U.}~\bibnamefont
  {Bovensiepen}},\ }\href {\doibase 10.1088/0953-8984/25/9/094003} {\bibfield
  {journal} {\bibinfo  {journal} {J. Phys.: Condens. Matter}\ }\textbf
  {\bibinfo {volume} {25}},\ \bibinfo {pages} {094003} (\bibinfo {year}
  {2013})}\BibitemShut {NoStop}%
\bibitem [{\citenamefont {Yang}\ \emph {et~al.}(2014)\citenamefont {Yang},
  \citenamefont {Rohde}, \citenamefont {Rohwer}, \citenamefont {Stange},
  \citenamefont {Hanff}, \citenamefont {Sohrt}, \citenamefont {Rettig},
  \citenamefont {Cort\'es}, \citenamefont {Chen}, \citenamefont {Feng},
  \citenamefont {Wolf}, \citenamefont {Kamble}, \citenamefont {Eremin},
  \citenamefont {Popmintchev}, \citenamefont {Murnane}, \citenamefont
  {Kapteyn}, \citenamefont {Kipp}, \citenamefont {Fink}, \citenamefont {Bauer},
  \citenamefont {Bovensiepen},\ and\ \citenamefont {Rossnagel}}]{Yang2014}%
  \BibitemOpen
  \bibfield  {author} {\bibinfo {author} {\bibfnamefont {L.~X.}\ \bibnamefont
  {Yang}}, \bibinfo {author} {\bibfnamefont {G.}~\bibnamefont {Rohde}},
  \bibinfo {author} {\bibfnamefont {T.}~\bibnamefont {Rohwer}}, \bibinfo
  {author} {\bibfnamefont {A.}~\bibnamefont {Stange}}, \bibinfo {author}
  {\bibfnamefont {K.}~\bibnamefont {Hanff}}, \bibinfo {author} {\bibfnamefont
  {C.}~\bibnamefont {Sohrt}}, \bibinfo {author} {\bibfnamefont
  {L.}~\bibnamefont {Rettig}}, \bibinfo {author} {\bibfnamefont
  {R.}~\bibnamefont {Cort\'es}}, \bibinfo {author} {\bibfnamefont
  {F.}~\bibnamefont {Chen}}, \bibinfo {author} {\bibfnamefont {D.~L.}\
  \bibnamefont {Feng}}, \bibinfo {author} {\bibfnamefont {T.}~\bibnamefont
  {Wolf}}, \bibinfo {author} {\bibfnamefont {B.}~\bibnamefont {Kamble}},
  \bibinfo {author} {\bibfnamefont {I.}~\bibnamefont {Eremin}}, \bibinfo
  {author} {\bibfnamefont {T.}~\bibnamefont {Popmintchev}}, \bibinfo {author}
  {\bibfnamefont {M.~M.}\ \bibnamefont {Murnane}}, \bibinfo {author}
  {\bibfnamefont {H.~C.}\ \bibnamefont {Kapteyn}}, \bibinfo {author}
  {\bibfnamefont {L.}~\bibnamefont {Kipp}}, \bibinfo {author} {\bibfnamefont
  {J.}~\bibnamefont {Fink}}, \bibinfo {author} {\bibfnamefont {M.}~\bibnamefont
  {Bauer}}, \bibinfo {author} {\bibfnamefont {U.}~\bibnamefont {Bovensiepen}},
  \ and\ \bibinfo {author} {\bibfnamefont {K.}~\bibnamefont {Rossnagel}},\
  }\href {\doibase 10.1103/PhysRevLett.112.207001} {\bibfield  {journal}
  {\bibinfo  {journal} {Phys. Rev. Lett.}\ }\textbf {\bibinfo {volume} {112}},\
  \bibinfo {pages} {207001} (\bibinfo {year} {2014})}\BibitemShut {NoStop}%
\bibitem [{\citenamefont {Rameau}\ \emph {et~al.}(2014)\citenamefont {Rameau},
  \citenamefont {Freutel}, \citenamefont {Rettig}, \citenamefont {Avigo},
  \citenamefont {Ligges}, \citenamefont {Yoshida}, \citenamefont {Eisaki},
  \citenamefont {Schneeloch}, \citenamefont {Zhong}, \citenamefont {Xu},
  \citenamefont {Gu}, \citenamefont {Johnson},\ and\ \citenamefont
  {Bovensiepen}}]{Rameau2014}%
  \BibitemOpen
  \bibfield  {author} {\bibinfo {author} {\bibfnamefont {J.~D.}\ \bibnamefont
  {Rameau}}, \bibinfo {author} {\bibfnamefont {S.}~\bibnamefont {Freutel}},
  \bibinfo {author} {\bibfnamefont {L.}~\bibnamefont {Rettig}}, \bibinfo
  {author} {\bibfnamefont {I.}~\bibnamefont {Avigo}}, \bibinfo {author}
  {\bibfnamefont {M.}~\bibnamefont {Ligges}}, \bibinfo {author} {\bibfnamefont
  {Y.}~\bibnamefont {Yoshida}}, \bibinfo {author} {\bibfnamefont
  {H.}~\bibnamefont {Eisaki}}, \bibinfo {author} {\bibfnamefont
  {J.}~\bibnamefont {Schneeloch}}, \bibinfo {author} {\bibfnamefont {R.~D.}\
  \bibnamefont {Zhong}}, \bibinfo {author} {\bibfnamefont {Z.~J.}\ \bibnamefont
  {Xu}}, \bibinfo {author} {\bibfnamefont {G.~D.}\ \bibnamefont {Gu}}, \bibinfo
  {author} {\bibfnamefont {P.~D.}\ \bibnamefont {Johnson}}, \ and\ \bibinfo
  {author} {\bibfnamefont {U.}~\bibnamefont {Bovensiepen}},\ }\href {\doibase
  10.1103/PhysRevB.89.115115} {\bibfield  {journal} {\bibinfo  {journal} {Phys.
  Rev. B}\ }\textbf {\bibinfo {volume} {89}},\ \bibinfo {pages} {115115}
  (\bibinfo {year} {2014})}\BibitemShut {NoStop}%
\bibitem [{\citenamefont {Smallwood}\ \emph {et~al.}(2014)\citenamefont
  {Smallwood}, \citenamefont {Zhang}, \citenamefont {Miller}, \citenamefont
  {Jozwiak}, \citenamefont {Eisaki}, \citenamefont {Lee},\ and\ \citenamefont
  {Lanzara}}]{Smallwood2014}%
  \BibitemOpen
  \bibfield  {author} {\bibinfo {author} {\bibfnamefont {C.~L.}\ \bibnamefont
  {Smallwood}}, \bibinfo {author} {\bibfnamefont {W.}~\bibnamefont {Zhang}},
  \bibinfo {author} {\bibfnamefont {T.~L.}\ \bibnamefont {Miller}}, \bibinfo
  {author} {\bibfnamefont {C.}~\bibnamefont {Jozwiak}}, \bibinfo {author}
  {\bibfnamefont {H.}~\bibnamefont {Eisaki}}, \bibinfo {author} {\bibfnamefont
  {D.-H.}\ \bibnamefont {Lee}}, \ and\ \bibinfo {author} {\bibfnamefont
  {A.}~\bibnamefont {Lanzara}},\ }\href {\doibase 10.1103/PhysRevB.89.115126}
  {\bibfield  {journal} {\bibinfo  {journal} {Phys. Rev. B}\ }\textbf {\bibinfo
  {volume} {89}},\ \bibinfo {pages} {115126} (\bibinfo {year}
  {2014})}\BibitemShut {NoStop}%
\bibitem [{\citenamefont {Clarke}(1972)}]{Clarke1972}%
  \BibitemOpen
  \bibfield  {author} {\bibinfo {author} {\bibfnamefont {J.}~\bibnamefont
  {Clarke}},\ }\href {\doibase 10.1103/PhysRevLett.28.1363} {\bibfield
  {journal} {\bibinfo  {journal} {Phys. Rev. Lett.}\ }\textbf {\bibinfo
  {volume} {28}},\ \bibinfo {pages} {1363} (\bibinfo {year}
  {1972})}\BibitemShut {NoStop}%
\bibitem [{\citenamefont {Tinkham}\ and\ \citenamefont
  {Clarke}(1972)}]{Tinkham1972}%
  \BibitemOpen
  \bibfield  {author} {\bibinfo {author} {\bibfnamefont {M.}~\bibnamefont
  {Tinkham}}\ and\ \bibinfo {author} {\bibfnamefont {J.}~\bibnamefont
  {Clarke}},\ }\href {\doibase 10.1103/PhysRevLett.28.1366} {\bibfield
  {journal} {\bibinfo  {journal} {Phys. Rev. Lett.}\ }\textbf {\bibinfo
  {volume} {28}},\ \bibinfo {pages} {1366} (\bibinfo {year}
  {1972})}\BibitemShut {NoStop}%
\bibitem [{\citenamefont {Smallwood}\ \emph
  {et~al.}(2012{\natexlab{a}})\citenamefont {Smallwood}, \citenamefont
  {Hinton}, \citenamefont {Jozwiak}, \citenamefont {Zhang}, \citenamefont
  {Koralek}, \citenamefont {Eisaki}, \citenamefont {Lee}, \citenamefont
  {Orenstein},\ and\ \citenamefont {Lanzara}}]{Smallwood2012}%
  \BibitemOpen
  \bibfield  {author} {\bibinfo {author} {\bibfnamefont {C.~L.}\ \bibnamefont
  {Smallwood}}, \bibinfo {author} {\bibfnamefont {J.~P.}\ \bibnamefont
  {Hinton}}, \bibinfo {author} {\bibfnamefont {C.}~\bibnamefont {Jozwiak}},
  \bibinfo {author} {\bibfnamefont {W.~T.}\ \bibnamefont {Zhang}}, \bibinfo
  {author} {\bibfnamefont {J.~D.}\ \bibnamefont {Koralek}}, \bibinfo {author}
  {\bibfnamefont {H.}~\bibnamefont {Eisaki}}, \bibinfo {author} {\bibfnamefont
  {D.~H.}\ \bibnamefont {Lee}}, \bibinfo {author} {\bibfnamefont
  {J.}~\bibnamefont {Orenstein}}, \ and\ \bibinfo {author} {\bibfnamefont
  {A.}~\bibnamefont {Lanzara}},\ }\href {\doibase 10.1126/science.1217423}
  {\bibfield  {journal} {\bibinfo  {journal} {Science}\ }\textbf {\bibinfo
  {volume} {336}},\ \bibinfo {pages} {1137} (\bibinfo {year}
  {2012}{\natexlab{a}})}\BibitemShut {NoStop}%
\bibitem [{\citenamefont {Zhang}\ \emph {et~al.}(2014)\citenamefont {Zhang},
  \citenamefont {Hwang}, \citenamefont {Smallwood}, \citenamefont {Miller},
  \citenamefont {Affeldt}, \citenamefont {Kurashima}, \citenamefont {Jozwiak},
  \citenamefont {Eisaki}, \citenamefont {Adachi}, \citenamefont {Koike},
  \citenamefont {Lee},\ and\ \citenamefont {Lanzara}}]{Zhang2014}%
  \BibitemOpen
  \bibfield  {author} {\bibinfo {author} {\bibfnamefont {W.}~\bibnamefont
  {Zhang}}, \bibinfo {author} {\bibfnamefont {C.}~\bibnamefont {Hwang}},
  \bibinfo {author} {\bibfnamefont {C.~L.}\ \bibnamefont {Smallwood}}, \bibinfo
  {author} {\bibfnamefont {T.~L.}\ \bibnamefont {Miller}}, \bibinfo {author}
  {\bibfnamefont {G.}~\bibnamefont {Affeldt}}, \bibinfo {author} {\bibfnamefont
  {K.}~\bibnamefont {Kurashima}}, \bibinfo {author} {\bibfnamefont
  {C.}~\bibnamefont {Jozwiak}}, \bibinfo {author} {\bibfnamefont
  {H.}~\bibnamefont {Eisaki}}, \bibinfo {author} {\bibfnamefont
  {T.}~\bibnamefont {Adachi}}, \bibinfo {author} {\bibfnamefont
  {Y.}~\bibnamefont {Koike}}, \bibinfo {author} {\bibfnamefont {D.-H.}\
  \bibnamefont {Lee}}, \ and\ \bibinfo {author} {\bibfnamefont
  {A.}~\bibnamefont {Lanzara}},\ }\href {\doibase 10.1038/ncomms5959}
  {\bibfield  {journal} {\bibinfo  {journal} {Nat. Commun.}\ }\textbf {\bibinfo
  {volume} {5}},\ \bibinfo {pages} {4959} (\bibinfo {year} {2014})}\BibitemShut
  {NoStop}%
\bibitem [{\citenamefont {Smallwood}\ \emph
  {et~al.}(2012{\natexlab{b}})\citenamefont {Smallwood}, \citenamefont
  {Jozwiak}, \citenamefont {Zhang},\ and\ \citenamefont
  {Lanzara}}]{Smallwood2012RSI}%
  \BibitemOpen
  \bibfield  {author} {\bibinfo {author} {\bibfnamefont {C.~L.}\ \bibnamefont
  {Smallwood}}, \bibinfo {author} {\bibfnamefont {C.}~\bibnamefont {Jozwiak}},
  \bibinfo {author} {\bibfnamefont {W.}~\bibnamefont {Zhang}}, \ and\ \bibinfo
  {author} {\bibfnamefont {A.}~\bibnamefont {Lanzara}},\ }\href {\doibase
  10.1063/1.4772070} {\bibfield  {journal} {\bibinfo  {journal} {Rev. Sci.
  Instrum.}\ }\textbf {\bibinfo {volume} {83}},\ \bibinfo {pages} {123904}
  (\bibinfo {year} {2012}{\natexlab{b}})}\BibitemShut {NoStop}%
\bibitem [{\citenamefont {Perfetti}\ \emph {et~al.}(2007)\citenamefont
  {Perfetti}, \citenamefont {Loukakos}, \citenamefont {Lisowski}, \citenamefont
  {Bovensiepen}, \citenamefont {Eisaki},\ and\ \citenamefont
  {Wolf}}]{Perfetti2007}%
  \BibitemOpen
  \bibfield  {author} {\bibinfo {author} {\bibfnamefont {L.}~\bibnamefont
  {Perfetti}}, \bibinfo {author} {\bibfnamefont {P.~A.}\ \bibnamefont
  {Loukakos}}, \bibinfo {author} {\bibfnamefont {M.}~\bibnamefont {Lisowski}},
  \bibinfo {author} {\bibfnamefont {U.}~\bibnamefont {Bovensiepen}}, \bibinfo
  {author} {\bibfnamefont {H.}~\bibnamefont {Eisaki}}, \ and\ \bibinfo {author}
  {\bibfnamefont {M.}~\bibnamefont {Wolf}},\ }\href {\doibase
  10.1103/PhysRevLett.99.197001} {\bibfield  {journal} {\bibinfo  {journal}
  {Phys. Rev. Lett.}\ }\textbf {\bibinfo {volume} {99}},\ \bibinfo {pages}
  {197001} (\bibinfo {year} {2007})}\BibitemShut {NoStop}%
\bibitem [{\citenamefont {Graf}\ \emph {et~al.}(2011)\citenamefont {Graf},
  \citenamefont {Jozwiak}, \citenamefont {Smallwood}, \citenamefont {Eisaki},
  \citenamefont {Kaindl}, \citenamefont {Lee},\ and\ \citenamefont
  {Lanzara}}]{Graf2011}%
  \BibitemOpen
  \bibfield  {author} {\bibinfo {author} {\bibfnamefont {J.}~\bibnamefont
  {Graf}}, \bibinfo {author} {\bibfnamefont {C.}~\bibnamefont {Jozwiak}},
  \bibinfo {author} {\bibfnamefont {C.~L.}\ \bibnamefont {Smallwood}}, \bibinfo
  {author} {\bibfnamefont {H.}~\bibnamefont {Eisaki}}, \bibinfo {author}
  {\bibfnamefont {R.~A.}\ \bibnamefont {Kaindl}}, \bibinfo {author}
  {\bibfnamefont {D.-H.}\ \bibnamefont {Lee}}, \ and\ \bibinfo {author}
  {\bibfnamefont {A.}~\bibnamefont {Lanzara}},\ }\href {\doibase
  10.1038/nphys2027} {\bibfield  {journal} {\bibinfo  {journal} {Nat. Phys.}\
  }\textbf {\bibinfo {volume} {7}},\ \bibinfo {pages} {805} (\bibinfo {year}
  {2011})}\BibitemShut {NoStop}%
\bibitem [{\citenamefont {Reber}\ \emph {et~al.}(2014)\citenamefont {Reber},
  \citenamefont {Plumb}, \citenamefont {Waugh},\ and\ \citenamefont
  {Dessau}}]{Reber2014}%
  \BibitemOpen
  \bibfield  {author} {\bibinfo {author} {\bibfnamefont {T.~J.}\ \bibnamefont
  {Reber}}, \bibinfo {author} {\bibfnamefont {N.~C.}\ \bibnamefont {Plumb}},
  \bibinfo {author} {\bibfnamefont {J.~A.}\ \bibnamefont {Waugh}}, \ and\
  \bibinfo {author} {\bibfnamefont {D.~S.}\ \bibnamefont {Dessau}},\ }\href
  {\doibase 10.1063/1.4870283} {\bibfield  {journal} {\bibinfo  {journal}
  {Review of Scientific Instruments}\ }\textbf {\bibinfo {volume} {85}},\
  \bibinfo {pages} {043907} (\bibinfo {year} {2014})}\BibitemShut {NoStop}%
\bibitem [{\citenamefont {Zhou}\ \emph {et~al.}(2005)\citenamefont {Zhou},
  \citenamefont {Wannberg}, \citenamefont {Yang}, \citenamefont {Brouet},
  \citenamefont {Sun}, \citenamefont {Douglas}, \citenamefont {Dessau},
  \citenamefont {Hussain},\ and\ \citenamefont {Shen}}]{Zhou2005}%
  \BibitemOpen
  \bibfield  {author} {\bibinfo {author} {\bibfnamefont {X.}~\bibnamefont
  {Zhou}}, \bibinfo {author} {\bibfnamefont {B.}~\bibnamefont {Wannberg}},
  \bibinfo {author} {\bibfnamefont {W.}~\bibnamefont {Yang}}, \bibinfo {author}
  {\bibfnamefont {V.}~\bibnamefont {Brouet}}, \bibinfo {author} {\bibfnamefont
  {Z.}~\bibnamefont {Sun}}, \bibinfo {author} {\bibfnamefont {J.}~\bibnamefont
  {Douglas}}, \bibinfo {author} {\bibfnamefont {D.}~\bibnamefont {Dessau}},
  \bibinfo {author} {\bibfnamefont {Z.}~\bibnamefont {Hussain}}, \ and\
  \bibinfo {author} {\bibfnamefont {Z.-X.}\ \bibnamefont {Shen}},\ }\href
  {\doibase 10.1016/j.elspec.2004.08.004} {\bibfield  {journal} {\bibinfo
  {journal} {J. Electron Spectrosc. Relat. Phenom.}\ }\textbf {\bibinfo
  {volume} {142}},\ \bibinfo {pages} {27} (\bibinfo {year} {2005})}\BibitemShut
  {NoStop}%
\bibitem [{\citenamefont {Graf}\ \emph {et~al.}(2010)\citenamefont {Graf},
  \citenamefont {Hellmann}, \citenamefont {Jozwiak}, \citenamefont {Smallwood},
  \citenamefont {Hussain}, \citenamefont {Kaindl}, \citenamefont {Kipp},
  \citenamefont {Rossnagel},\ and\ \citenamefont {Lanzara}}]{Graf2010}%
  \BibitemOpen
  \bibfield  {author} {\bibinfo {author} {\bibfnamefont {J.}~\bibnamefont
  {Graf}}, \bibinfo {author} {\bibfnamefont {S.}~\bibnamefont {Hellmann}},
  \bibinfo {author} {\bibfnamefont {C.}~\bibnamefont {Jozwiak}}, \bibinfo
  {author} {\bibfnamefont {C.~L.}\ \bibnamefont {Smallwood}}, \bibinfo {author}
  {\bibfnamefont {Z.}~\bibnamefont {Hussain}}, \bibinfo {author} {\bibfnamefont
  {R.~A.}\ \bibnamefont {Kaindl}}, \bibinfo {author} {\bibfnamefont
  {L.}~\bibnamefont {Kipp}}, \bibinfo {author} {\bibfnamefont {K.}~\bibnamefont
  {Rossnagel}}, \ and\ \bibinfo {author} {\bibfnamefont {A.}~\bibnamefont
  {Lanzara}},\ }\href {\doibase 10.1063/1.3273487} {\bibfield  {journal}
  {\bibinfo  {journal} {Journal of Applied Physics}\ }\textbf {\bibinfo
  {volume} {107}},\ \bibinfo {eid} {014912} (\bibinfo {year}
  {2010})}\BibitemShut {NoStop}%
\bibitem [{\citenamefont {LaShell}\ \emph {et~al.}(2000)\citenamefont
  {LaShell}, \citenamefont {Jensen},\ and\ \citenamefont
  {Balasubramanian}}]{LaShell2000}%
  \BibitemOpen
  \bibfield  {author} {\bibinfo {author} {\bibfnamefont {S.}~\bibnamefont
  {LaShell}}, \bibinfo {author} {\bibfnamefont {E.}~\bibnamefont {Jensen}}, \
  and\ \bibinfo {author} {\bibfnamefont {T.}~\bibnamefont {Balasubramanian}},\
  }\href {\doibase 10.1103/PhysRevB.61.2371} {\bibfield  {journal} {\bibinfo
  {journal} {Phys. Rev. B}\ }\textbf {\bibinfo {volume} {61}},\ \bibinfo
  {pages} {2371} (\bibinfo {year} {2000})}\BibitemShut {NoStop}%
\bibitem [{\citenamefont {Levy}\ \emph {et~al.}(2014)\citenamefont {Levy},
  \citenamefont {Nettke}, \citenamefont {Ludbrook}, \citenamefont {Veenstra},\
  and\ \citenamefont {Damascelli}}]{Levy2014}%
  \BibitemOpen
  \bibfield  {author} {\bibinfo {author} {\bibfnamefont {G.}~\bibnamefont
  {Levy}}, \bibinfo {author} {\bibfnamefont {W.}~\bibnamefont {Nettke}},
  \bibinfo {author} {\bibfnamefont {B.~M.}\ \bibnamefont {Ludbrook}}, \bibinfo
  {author} {\bibfnamefont {C.~N.}\ \bibnamefont {Veenstra}}, \ and\ \bibinfo
  {author} {\bibfnamefont {A.}~\bibnamefont {Damascelli}},\ }\href {\doibase
  10.1103/PhysRevB.90.045150} {\bibfield  {journal} {\bibinfo  {journal} {Phys.
  Rev. B}\ }\textbf {\bibinfo {volume} {90}},\ \bibinfo {pages} {045150}
  (\bibinfo {year} {2014})}\BibitemShut {NoStop}%
\bibitem [{\citenamefont {Seah}\ and\ \citenamefont {Dench}(1979)}]{Seah1979}%
  \BibitemOpen
  \bibfield  {author} {\bibinfo {author} {\bibfnamefont {M.~P.}\ \bibnamefont
  {Seah}}\ and\ \bibinfo {author} {\bibfnamefont {W.~A.}\ \bibnamefont
  {Dench}},\ }\href {\doibase 10.1002/sia.740010103} {\bibfield  {journal}
  {\bibinfo  {journal} {Surf. Interface Anal.}\ }\textbf {\bibinfo {volume}
  {1}},\ \bibinfo {pages} {2} (\bibinfo {year} {1979})}\BibitemShut {NoStop}%
\bibitem [{\citenamefont {Smallwood}\ \emph {et~al.}(2015)\citenamefont
  {Smallwood}, \citenamefont {Zhang}, \citenamefont {Miller}, \citenamefont
  {Affeldt}, \citenamefont {Kurashima}, \citenamefont {Jozwiak}, \citenamefont
  {Noji}, \citenamefont {Koike}, \citenamefont {Eisaki}, \citenamefont {Lee},
  \citenamefont {Kaindl},\ and\ \citenamefont {Lanzara}}]{Smallwood2015}%
  \BibitemOpen
  \bibfield  {author} {\bibinfo {author} {\bibfnamefont {C.~L.}\ \bibnamefont
  {Smallwood}}, \bibinfo {author} {\bibfnamefont {W.}~\bibnamefont {Zhang}},
  \bibinfo {author} {\bibfnamefont {T.~L.}\ \bibnamefont {Miller}}, \bibinfo
  {author} {\bibfnamefont {G.}~\bibnamefont {Affeldt}}, \bibinfo {author}
  {\bibfnamefont {K.}~\bibnamefont {Kurashima}}, \bibinfo {author}
  {\bibfnamefont {C.}~\bibnamefont {Jozwiak}}, \bibinfo {author} {\bibfnamefont
  {T.}~\bibnamefont {Noji}}, \bibinfo {author} {\bibfnamefont {Y.}~\bibnamefont
  {Koike}}, \bibinfo {author} {\bibfnamefont {H.}~\bibnamefont {Eisaki}},
  \bibinfo {author} {\bibfnamefont {D.-H.}\ \bibnamefont {Lee}}, \bibinfo
  {author} {\bibfnamefont {R.~A.}\ \bibnamefont {Kaindl}}, \ and\ \bibinfo
  {author} {\bibfnamefont {A.}~\bibnamefont {Lanzara}},\ }\href
  {http://link.aps.org/doi/10.1103/PhysRevB.92.161102} {\bibfield  {journal}
  {\bibinfo  {journal} {Phys. Rev. B}\ }\textbf {\bibinfo {volume} {92}},\
  \bibinfo {pages} {161102} (\bibinfo {year} {2015})}\BibitemShut {NoStop}%
\bibitem [{\citenamefont {Norman}\ \emph {et~al.}(1995)\citenamefont {Norman},
  \citenamefont {Randeria}, \citenamefont {Ding},\ and\ \citenamefont
  {Campuzano}}]{Norman1995}%
  \BibitemOpen
  \bibfield  {author} {\bibinfo {author} {\bibfnamefont {M.~R.}\ \bibnamefont
  {Norman}}, \bibinfo {author} {\bibfnamefont {M.}~\bibnamefont {Randeria}},
  \bibinfo {author} {\bibfnamefont {H.}~\bibnamefont {Ding}}, \ and\ \bibinfo
  {author} {\bibfnamefont {J.~C.}\ \bibnamefont {Campuzano}},\ }\href {\doibase
  10.1103/PhysRevB.52.615} {\bibfield  {journal} {\bibinfo  {journal} {Phys.
  Rev. B}\ }\textbf {\bibinfo {volume} {52}},\ \bibinfo {pages} {615} (\bibinfo
  {year} {1995})}\BibitemShut {NoStop}%
\bibitem [{\citenamefont {Yang}\ \emph {et~al.}(2008)\citenamefont {Yang},
  \citenamefont {Rameau}, \citenamefont {Johnson}, \citenamefont {Valla},
  \citenamefont {Tsvelik},\ and\ \citenamefont {Gu}}]{Yang2008}%
  \BibitemOpen
  \bibfield  {author} {\bibinfo {author} {\bibfnamefont {H.-B.}\ \bibnamefont
  {Yang}}, \bibinfo {author} {\bibfnamefont {J.~D.}\ \bibnamefont {Rameau}},
  \bibinfo {author} {\bibfnamefont {P.~D.}\ \bibnamefont {Johnson}}, \bibinfo
  {author} {\bibfnamefont {T.}~\bibnamefont {Valla}}, \bibinfo {author}
  {\bibfnamefont {A.}~\bibnamefont {Tsvelik}}, \ and\ \bibinfo {author}
  {\bibfnamefont {G.~D.}\ \bibnamefont {Gu}},\ }\href {\doibase
  10.1038/nature07400} {\bibfield  {journal} {\bibinfo  {journal} {Nature}\
  }\textbf {\bibinfo {volume} {456}},\ \bibinfo {pages} {77} (\bibinfo {year}
  {2008})}\BibitemShut {NoStop}%
\bibitem [{\citenamefont {Hashimoto}\ \emph {et~al.}(2010)\citenamefont
  {Hashimoto}, \citenamefont {He}, \citenamefont {Tanaka}, \citenamefont
  {Testaud}, \citenamefont {Meevasana}, \citenamefont {Moore}, \citenamefont
  {Lu}, \citenamefont {Yao}, \citenamefont {Yoshida}, \citenamefont {Eisaki},
  \citenamefont {Devereaux}, \citenamefont {Hussain},\ and\ \citenamefont
  {Shen}}]{Hashimoto2010}%
  \BibitemOpen
  \bibfield  {author} {\bibinfo {author} {\bibfnamefont {M.}~\bibnamefont
  {Hashimoto}}, \bibinfo {author} {\bibfnamefont {R.-H.}\ \bibnamefont {He}},
  \bibinfo {author} {\bibfnamefont {K.}~\bibnamefont {Tanaka}}, \bibinfo
  {author} {\bibfnamefont {J.-P.}\ \bibnamefont {Testaud}}, \bibinfo {author}
  {\bibfnamefont {W.}~\bibnamefont {Meevasana}}, \bibinfo {author}
  {\bibfnamefont {R.~G.}\ \bibnamefont {Moore}}, \bibinfo {author}
  {\bibfnamefont {D.}~\bibnamefont {Lu}}, \bibinfo {author} {\bibfnamefont
  {H.}~\bibnamefont {Yao}}, \bibinfo {author} {\bibfnamefont {Y.}~\bibnamefont
  {Yoshida}}, \bibinfo {author} {\bibfnamefont {H.}~\bibnamefont {Eisaki}},
  \bibinfo {author} {\bibfnamefont {T.~P.}\ \bibnamefont {Devereaux}}, \bibinfo
  {author} {\bibfnamefont {Z.}~\bibnamefont {Hussain}}, \ and\ \bibinfo
  {author} {\bibfnamefont {Z.-X.}\ \bibnamefont {Shen}},\ }\href {\doibase
  10.1038/nphys1632} {\bibfield  {journal} {\bibinfo  {journal} {Nat. Phys.}\
  }\textbf {\bibinfo {volume} {6}},\ \bibinfo {pages} {414} (\bibinfo {year}
  {2010})}\BibitemShut {NoStop}%
\bibitem [{\citenamefont {Sakai}\ \emph {et~al.}(2013)\citenamefont {Sakai},
  \citenamefont {Blanc}, \citenamefont {Civelli}, \citenamefont {Gallais},
  \citenamefont {Cazayous}, \citenamefont {M\'easson}, \citenamefont {Wen},
  \citenamefont {Xu}, \citenamefont {Gu}, \citenamefont {Sangiovanni},
  \citenamefont {Motome}, \citenamefont {Held}, \citenamefont {Sacuto},
  \citenamefont {Georges},\ and\ \citenamefont {Imada}}]{Sakai2013}%
  \BibitemOpen
  \bibfield  {author} {\bibinfo {author} {\bibfnamefont {S.}~\bibnamefont
  {Sakai}}, \bibinfo {author} {\bibfnamefont {S.}~\bibnamefont {Blanc}},
  \bibinfo {author} {\bibfnamefont {M.}~\bibnamefont {Civelli}}, \bibinfo
  {author} {\bibfnamefont {Y.}~\bibnamefont {Gallais}}, \bibinfo {author}
  {\bibfnamefont {M.}~\bibnamefont {Cazayous}}, \bibinfo {author}
  {\bibfnamefont {M.-A.}\ \bibnamefont {M\'easson}}, \bibinfo {author}
  {\bibfnamefont {J.~S.}\ \bibnamefont {Wen}}, \bibinfo {author} {\bibfnamefont
  {Z.~J.}\ \bibnamefont {Xu}}, \bibinfo {author} {\bibfnamefont {G.~D.}\
  \bibnamefont {Gu}}, \bibinfo {author} {\bibfnamefont {G.}~\bibnamefont
  {Sangiovanni}}, \bibinfo {author} {\bibfnamefont {Y.}~\bibnamefont {Motome}},
  \bibinfo {author} {\bibfnamefont {K.}~\bibnamefont {Held}}, \bibinfo {author}
  {\bibfnamefont {A.}~\bibnamefont {Sacuto}}, \bibinfo {author} {\bibfnamefont
  {A.}~\bibnamefont {Georges}}, \ and\ \bibinfo {author} {\bibfnamefont
  {M.}~\bibnamefont {Imada}},\ }\href {\doibase 10.1103/PhysRevLett.111.107001}
  {\bibfield  {journal} {\bibinfo  {journal} {Phys. Rev. Lett.}\ }\textbf
  {\bibinfo {volume} {111}},\ \bibinfo {pages} {107001} (\bibinfo {year}
  {2013})}\BibitemShut {NoStop}%
\bibitem [{\citenamefont {Rice}\ \emph {et~al.}(2012)\citenamefont {Rice},
  \citenamefont {Yang},\ and\ \citenamefont {Zhang}}]{Rice2012}%
  \BibitemOpen
  \bibfield  {author} {\bibinfo {author} {\bibfnamefont {T.~M.}\ \bibnamefont
  {Rice}}, \bibinfo {author} {\bibfnamefont {K.-Y.}\ \bibnamefont {Yang}}, \
  and\ \bibinfo {author} {\bibfnamefont {F.~C.}\ \bibnamefont {Zhang}},\ }\href
  {\doibase 10.1088/0034-4885/75/1/016502} {\bibfield  {journal} {\bibinfo
  {journal} {Reports on Progress in Physics}\ }\textbf {\bibinfo {volume}
  {75}},\ \bibinfo {pages} {016502} (\bibinfo {year} {2012})}\BibitemShut
  {NoStop}%
\bibitem [{\citenamefont {Tinkham}(1996)}]{Tinkham1996}%
  \BibitemOpen
  \bibfield  {author} {\bibinfo {author} {\bibfnamefont {M.}~\bibnamefont
  {Tinkham}},\ }\href@noop {} {\emph {\bibinfo {title} {Introduction to
  Superconductivity}}}\ (\bibinfo  {publisher} {McGraw-Hill, New York},\
  \bibinfo {year} {1996})\BibitemShut {NoStop}%
\bibitem [{\citenamefont {Van~der Marel}(1990)}]{Marel1990}%
  \BibitemOpen
  \bibfield  {author} {\bibinfo {author} {\bibfnamefont {D.}~\bibnamefont
  {Van~der Marel}},\ }\href {\doibase 10.1016/0921-4534(90)90429-I} {\bibfield
  {journal} {\bibinfo  {journal} {Physica C: Superconductivity}\ }\textbf
  {\bibinfo {volume} {165}},\ \bibinfo {pages} {35} (\bibinfo {year}
  {1990})}\BibitemShut {NoStop}%
\bibitem [{\citenamefont {Vishik}\ \emph {et~al.}(2012)\citenamefont {Vishik},
  \citenamefont {Hashimoto}, \citenamefont {He}, \citenamefont {Lee},
  \citenamefont {Schmitt}, \citenamefont {Lu}, \citenamefont {Moore},
  \citenamefont {Zhang}, \citenamefont {Meevasana}, \citenamefont {Sasagawa},
  \citenamefont {Uchida}, \citenamefont {Fujita}, \citenamefont {Ishida},
  \citenamefont {Ishikado}, \citenamefont {Yoshida}, \citenamefont {Eisaki},
  \citenamefont {Hussain}, \citenamefont {Devereaux},\ and\ \citenamefont
  {Shen}}]{Vishik2012}%
  \BibitemOpen
  \bibfield  {author} {\bibinfo {author} {\bibfnamefont {I.~M.}\ \bibnamefont
  {Vishik}}, \bibinfo {author} {\bibfnamefont {M.}~\bibnamefont {Hashimoto}},
  \bibinfo {author} {\bibfnamefont {R.-H.}\ \bibnamefont {He}}, \bibinfo
  {author} {\bibfnamefont {W.-S.}\ \bibnamefont {Lee}}, \bibinfo {author}
  {\bibfnamefont {F.}~\bibnamefont {Schmitt}}, \bibinfo {author} {\bibfnamefont
  {D.}~\bibnamefont {Lu}}, \bibinfo {author} {\bibfnamefont {R.~G.}\
  \bibnamefont {Moore}}, \bibinfo {author} {\bibfnamefont {C.}~\bibnamefont
  {Zhang}}, \bibinfo {author} {\bibfnamefont {W.}~\bibnamefont {Meevasana}},
  \bibinfo {author} {\bibfnamefont {T.}~\bibnamefont {Sasagawa}}, \bibinfo
  {author} {\bibfnamefont {S.}~\bibnamefont {Uchida}}, \bibinfo {author}
  {\bibfnamefont {K.}~\bibnamefont {Fujita}}, \bibinfo {author} {\bibfnamefont
  {S.}~\bibnamefont {Ishida}}, \bibinfo {author} {\bibfnamefont
  {M.}~\bibnamefont {Ishikado}}, \bibinfo {author} {\bibfnamefont
  {Y.}~\bibnamefont {Yoshida}}, \bibinfo {author} {\bibfnamefont
  {H.}~\bibnamefont {Eisaki}}, \bibinfo {author} {\bibfnamefont
  {Z.}~\bibnamefont {Hussain}}, \bibinfo {author} {\bibfnamefont {T.~P.}\
  \bibnamefont {Devereaux}}, \ and\ \bibinfo {author} {\bibfnamefont {Z.-X.}\
  \bibnamefont {Shen}},\ }\href {\doibase 10.1073/pnas.1209471109} {\bibfield
  {journal} {\bibinfo  {journal} {Proc. Natl. Acad. Sci. USA}\ }\textbf
  {\bibinfo {volume} {109}},\ \bibinfo {pages} {18332} (\bibinfo {year}
  {2012})}\BibitemShut {NoStop}%
\bibitem [{\citenamefont {Pan}\ \emph {et~al.}(2001)\citenamefont {Pan},
  \citenamefont {O'Neal}, \citenamefont {Badzey}, \citenamefont {Chamon},
  \citenamefont {Ding}, \citenamefont {Engelbrecht}, \citenamefont {Wang},
  \citenamefont {Eisaki}, \citenamefont {Uchida}, \citenamefont {Gupta},
  \citenamefont {Ng}, \citenamefont {Hudson}, \citenamefont {Lang},\ and\
  \citenamefont {Davis}}]{Pan2001}%
  \BibitemOpen
  \bibfield  {author} {\bibinfo {author} {\bibfnamefont {S.~H.}\ \bibnamefont
  {Pan}}, \bibinfo {author} {\bibfnamefont {J.~P.}\ \bibnamefont {O'Neal}},
  \bibinfo {author} {\bibfnamefont {R.~L.}\ \bibnamefont {Badzey}}, \bibinfo
  {author} {\bibfnamefont {C.}~\bibnamefont {Chamon}}, \bibinfo {author}
  {\bibfnamefont {H.}~\bibnamefont {Ding}}, \bibinfo {author} {\bibfnamefont
  {J.~R.}\ \bibnamefont {Engelbrecht}}, \bibinfo {author} {\bibfnamefont
  {Z.}~\bibnamefont {Wang}}, \bibinfo {author} {\bibfnamefont {H.}~\bibnamefont
  {Eisaki}}, \bibinfo {author} {\bibfnamefont {S.}~\bibnamefont {Uchida}},
  \bibinfo {author} {\bibfnamefont {A.~K.}\ \bibnamefont {Gupta}}, \bibinfo
  {author} {\bibfnamefont {K.-W.}\ \bibnamefont {Ng}}, \bibinfo {author}
  {\bibfnamefont {E.~W.}\ \bibnamefont {Hudson}}, \bibinfo {author}
  {\bibfnamefont {K.~M.}\ \bibnamefont {Lang}}, \ and\ \bibinfo {author}
  {\bibfnamefont {J.~C.}\ \bibnamefont {Davis}},\ }\href {\doibase
  10.1038/35095012} {\bibfield  {journal} {\bibinfo  {journal} {Nature}\
  }\textbf {\bibinfo {volume} {413}},\ \bibinfo {pages} {282} (\bibinfo {year}
  {2001})}\BibitemShut {NoStop}%
\bibitem [{\citenamefont {Lang}\ \emph {et~al.}(2002)\citenamefont {Lang},
  \citenamefont {Madhavan}, \citenamefont {Hoffman}, \citenamefont {Hudson},
  \citenamefont {Eisaki}, \citenamefont {Uchida},\ and\ \citenamefont
  {Davis}}]{Lang2002}%
  \BibitemOpen
  \bibfield  {author} {\bibinfo {author} {\bibfnamefont {K.~M.}\ \bibnamefont
  {Lang}}, \bibinfo {author} {\bibfnamefont {V.}~\bibnamefont {Madhavan}},
  \bibinfo {author} {\bibfnamefont {J.~E.}\ \bibnamefont {Hoffman}}, \bibinfo
  {author} {\bibfnamefont {E.~W.}\ \bibnamefont {Hudson}}, \bibinfo {author}
  {\bibfnamefont {H.}~\bibnamefont {Eisaki}}, \bibinfo {author} {\bibfnamefont
  {S.}~\bibnamefont {Uchida}}, \ and\ \bibinfo {author} {\bibfnamefont {J.~C.}\
  \bibnamefont {Davis}},\ }\href {\doibase 10.1038/415412a} {\bibfield
  {journal} {\bibinfo  {journal} {Nature}\ }\textbf {\bibinfo {volume} {415}},\
  \bibinfo {pages} {412} (\bibinfo {year} {2002})}\BibitemShut {NoStop}%
\bibitem [{\citenamefont {Mouallem-Bahout}\ \emph {et~al.}(1994)\citenamefont
  {Mouallem-Bahout}, \citenamefont {Gaudé}, \citenamefont {Calvarin},
  \citenamefont {Gavarri},\ and\ \citenamefont {Carel}}]{Mouallem-Bahout1994}%
  \BibitemOpen
  \bibfield  {author} {\bibinfo {author} {\bibfnamefont {M.}~\bibnamefont
  {Mouallem-Bahout}}, \bibinfo {author} {\bibfnamefont {J.}~\bibnamefont
  {Gaudé}}, \bibinfo {author} {\bibfnamefont {G.}~\bibnamefont {Calvarin}},
  \bibinfo {author} {\bibfnamefont {J.-R.}\ \bibnamefont {Gavarri}}, \ and\
  \bibinfo {author} {\bibfnamefont {C.}~\bibnamefont {Carel}},\ }\href
  {\doibase 10.1016/0167-577X(94)90227-5} {\bibfield  {journal} {\bibinfo
  {journal} {Mater. Lett.}\ }\textbf {\bibinfo {volume} {18}},\ \bibinfo
  {pages} {181} (\bibinfo {year} {1994})}\BibitemShut {NoStop}%
\bibitem [{\citenamefont {Asahi}\ \emph {et~al.}(1997)\citenamefont {Asahi},
  \citenamefont {Suzuki}, \citenamefont {Nakamura}, \citenamefont {Takano},\
  and\ \citenamefont {Kobayashi}}]{Asahi1997}%
  \BibitemOpen
  \bibfield  {author} {\bibinfo {author} {\bibfnamefont {T.}~\bibnamefont
  {Asahi}}, \bibinfo {author} {\bibfnamefont {H.}~\bibnamefont {Suzuki}},
  \bibinfo {author} {\bibfnamefont {M.}~\bibnamefont {Nakamura}}, \bibinfo
  {author} {\bibfnamefont {H.}~\bibnamefont {Takano}}, \ and\ \bibinfo {author}
  {\bibfnamefont {J.}~\bibnamefont {Kobayashi}},\ }\href {\doibase
  10.1103/PhysRevB.55.9125} {\bibfield  {journal} {\bibinfo  {journal} {Phys.
  Rev. B}\ }\textbf {\bibinfo {volume} {55}},\ \bibinfo {pages} {9125}
  (\bibinfo {year} {1997})}\BibitemShut {NoStop}%
\end{thebibliography}%

\end{document}